\documentclass[aps, pra, twocolumn, amsmath, amssymb, superscriptaddress, nofootinbib]{revtex4-1}

\makeatletter
\def\p@subsection{}
\def\p@subsubsection{}
\makeatother

\usepackage[utf8]{inputenc}
\usepackage{graphicx}
\usepackage{units}
\usepackage{color}
\usepackage[pdftex, colorlinks=true, linkcolor=myblue, citecolor=myblue,
urlcolor=myblue]{hyperref}
\usepackage{times}
\usepackage{comment}
\usepackage{graphicx}
\usepackage{amsmath, calc}
\usepackage{mathrsfs}
\usepackage{amsfonts}
\usepackage{amssymb}
\usepackage{bm}
\usepackage{bbm}
\usepackage{color}
\usepackage{array}
\usepackage{units}
\usepackage{tabstackengine}
\stackMath

\usepackage{empheq} 
\usepackage{cases}

\definecolor{myblue}{rgb}{0,0,1}
\definecolor{myred}{rgb}{1,0,0}

\newcommand{\bra}[1]{\langle #1|}
\newcommand{\ket}[1]{|#1\rangle}

\DeclareMathOperator{\Tr}{Tr}

\begin{document}


\title{Energy storage in a continuous-variable quantum battery with nonlinear coupling}


\author{C. A. Downing}
\email{c.a.downing@exeter.ac.uk} 
\affiliation{Department of Physics and Astronomy, University of Exeter, Exeter EX4 4QL, United Kingdom}

\author{M. S. Ukhtary} 
\affiliation{Research Center for Quantum Physics, National Research and Innovation Agency (BRIN), South Tangerang 15314, Indonesia}


\date{\today}


\begin{abstract}
In the quantum world, the process of energy storage can be enhanced thanks to various non-classical phenomena. This inspiring fact suggests quantum batteries as plausible sources of power for future quantum devices, at least in principle. However, thermodynamically not all of the energy stored in a quantum battery is useful for doing work. By considering a class of models based upon quantum continuous variables, here we show how the maximum extractable energy from a bosonic quantum battery can be intimately related to Heisenberg’s uncertainty principle. We found that realizing minimum uncertainty essentially guarantees that all of the energy stored in a Gaussian quantum battery can be withdrawn and used to do work. For a standard system where the charger and battery are coupled linearly, this criterion is satisfied rather trivially. However, our theoretical results demonstrate that -- for a quantum battery with nonlinear coupling -- a state of minimum uncertainty can also be achieved nontrivially via the generation of quantum squeezing. We characterize the charging performance of our proposed continuous variable quantum batteries in detail, and we hope that our theory may be useful in the design of a new generation of efficient quantum batteries harnessing bosonic excitations, such as those built with photonic architectures.
\end{abstract}


\maketitle



\section{Introduction}

Quantum batteries are energy storage devices composed of fundamental building blocks (qubits, qutrits and so on)~\cite{Quach2023, Campaioli2023}, which aim to exploit quantum phenomena in order to improve their performance~\cite{Rossini2020, Andolina2024}. In the last few years, the first steps towards the experimental realization of viable quantum batteries have already been taken, with steady progress made in the quantum coherent control of such objects~\cite{Quach2022, Joshi2022, Hu2022, Zheng2023, Qu2023, Huang2023, Niu2024, Donelli2025}. Most notably, there is vast potential for building quantum batteries from organic microcavities~\cite{Hymas2025}, and their credible application as an energy source for quantum computers has recently been suggested~\cite{Kurman2025}.

\begin{figure*}[tb]
 \includegraphics[width=1.0\linewidth]{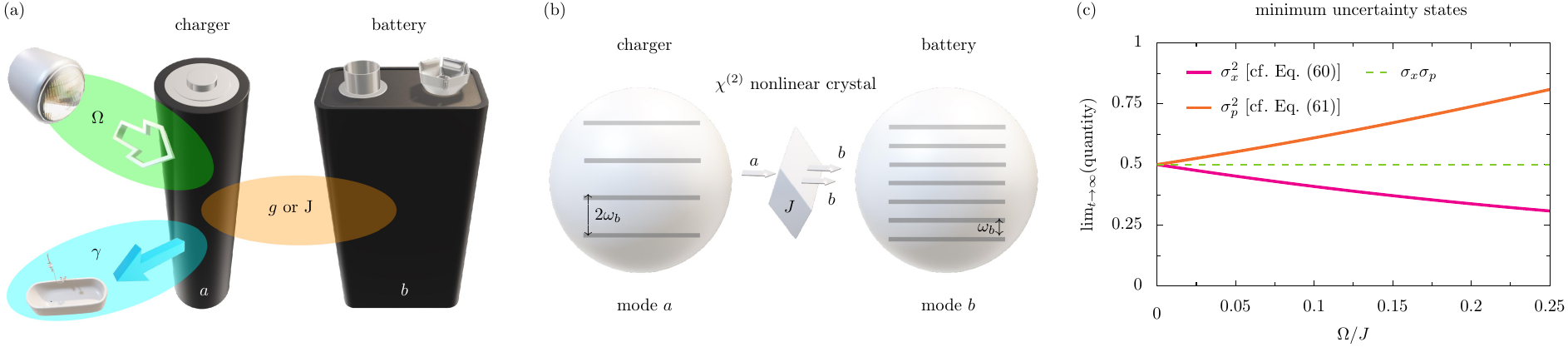}
 \caption{\textbf{The bosonic quantum battery as a driven-dissipative system.} Panel (a): the general setup of the two-component quantum battery, complete with its charger and battery pieces (modes $a$ and $b$). The charger is driven by laser light with an amplitude $\Omega$ (green region), and it suffers from dissipation at the rate $\gamma$ into its surrounding heat bath (blue region). Two flavours of interaction (orange region) are considered: either (i) linear coupling (of strength $g$) or (ii) nonlinear coupling (of strength $J$). These couplings enable excitation transfer from the time $t = 0$ from the charger mode $a$ into the battery mode $b$, where the energy is eventually stored in once the interaction is turned off at $t = T$. Panel (b): The nonlinear quantum battery is modelled as two quantum harmonic oscillators interacting via a nonlinear coupling of strength $J$, which is mediated by (for example) a $\chi^{(2)}$ nonlinear crystal. The emission of one photon of frequency $2\omega_b$ from the charger (mode $a$) leads to two photons of frequency $\omega_b$ impinging on the battery (mode $b$). Panel (c): the quadrature variances $\sigma_x^2$ and $\sigma_p^2$ in the steady state ($t \to \infty$) as a function of the dimensionless ratio $\Omega/J$ for the nonlinear quantum battery [cf. Eq.~\eqref{eq:xfgxsdfdfg} and Eq.~\eqref{eq:xfgxsdfsdddfg}]. Dashed green line: guide for the eye at the minimum uncertainty.}
 \label{squeez}
\end{figure*}

Thus far, the majority of theoretical attention on quantum batteries has been focussed on proposing devices working with discrete variables~\cite{Zinner2019, Saguia2020, Kamin2020, Kamin2021, Santos2021, Arjmandi2022, Arjmandi2023, Mojaveri2024, Wang2025}. Much less work has been conducted on batteries exploiting quantum continuous variables~\cite{Ahmadi2023, Gangwar2024, Ahmadi2024}, despite some intriguing advantages (including a lack of saturation due to their infinite dimension, and potentially attractive experimental realizations in photonic architectures -- such as that assembled in Ref.~\cite{Qu2023}). To aid the filling of this gap, here we consider a class of bosonic quantum batteries in an open quantum systems approach. We suppose that these driven-dissipative quantum batteries are built from two components, namely a charger mode $a$ and a battery mode $b$, with some coupling between them [as sketched in Fig.~\ref{squeez}~(a)]. During the charging phase of this device, from the initial time $t = 0$ and until $t = T$, the coupling will be turned on. Outside of this temporal window the charger will be disconnected from the battery, where the energy will then be stored until it is needed.

Most notably, we reveal the importance of generating minimum uncertainty states for the thermodynamic performance of Gaussian quantum batteries, via a criterion which is intrinsically linked to Heisenberg’s uncertainty principle. As examples of this credo, after discussing linear charger--battery coupling we go on to consider the more interesting example of quantum battery with a nonlinearity~\cite{Andolina2024, Bhattacharyya2024}, which is represented pictorially in Fig.~\ref{squeez}~(b). Such a setup is shown to be a nontrivial example of a quantum battery which supports a minimum uncertainty state, whereby the phenomenon of quantum squeezing~\cite{Walls1983, Loudon1987} gives rise to the desired minimum uncertainty in a nontrivial manner [as already previewed in Fig.~\ref{squeez}~(c)] and hence a thermodynamically efficient energy storage device.
 
In what follows, we sketch out the theory of quantum batteries utilizing quantum continuous variables in Sec.~\ref{eq:Bosonic}, before consider two types of interaction: the (essentially classical) case of linear charger--battery coupling in Sec.~\ref{eq:linear}, and the arguably more intriguing case of nonlinear charger--battery coupling in Sec.~\ref{eq:Nonlinear}. We draw together some conclusions on the prospects for bosonic quantum batteries in Sec.~\ref{eq:Discussion}, while the Supplemental Material contains some auxiliary calculations~\cite{Supplemental}.

\section{Bosonic battery}
\label{eq:Bosonic}

Let us consider a quantum battery defined by quantum continuous variables~\cite{Friis2018, Centrone2023, Ukhtary2024, Konar2024}. The bosonic quantum fields are represented by a single pair of self-adjoint canonical operators labelled $\hat{x}$ and $\hat{p}$, which are subject to the canonical commutation relation $[\hat{x}, \hat{p}] = \mathrm{i}$ (since we take $\hbar = 1 $ here and throughout). For a quantum state $\rho$, the two-dimensional vector $\boldsymbol{r}$ of first moments and the $2 \times 2$ covariance matrix $\boldsymbol{\sigma}$ of second moments are together defined by~\cite{Ferraro2005, Olivares2012}
\begin{align}
\label{eq:coxspvdsv}
\boldsymbol{r} &= \begin{pmatrix}
\langle \hat{x} \rangle  \\
\langle \hat{p} \rangle 
\end{pmatrix}, \\
\boldsymbol{\sigma} &= \begin{pmatrix}
\langle \{ \hat{x} - \langle \hat{x} \rangle, \hat{x} - \langle \hat{x} \rangle \} \rangle & \langle \{ \hat{x} - \langle \hat{x} \rangle, \hat{p} - \langle \hat{p} \rangle \} \rangle \\
\langle \{ \hat{p} - \langle \hat{p} \rangle, \hat{x} - \langle \hat{x} \rangle \} \rangle & \langle \{ \hat{p} - \langle \hat{p} \rangle, \hat{p} - \langle \hat{p} \rangle \} \rangle
\end{pmatrix}, \label{eq:coxspvddfsfvsv}
\end{align}
with the anticommutator $\{A, B\} = AB + BA$, and where the expectation value of some operator $\hat{\mathcal{O}}$ (for a given quantum state $\rho$) is given by the trace relation $\langle \hat{\mathcal{O}} \rangle = \Tr{(\rho \hat{\mathcal{O}})}$. The covariance matrix $\boldsymbol{\sigma}$ in Eq.~\eqref{eq:coxspvddfsfvsv} is equivalently determinable using the two quadrature variances $\sigma_x^2$ and $\sigma_p^2$ and the coherence $\xi$ as follows
\begin{align}
\boldsymbol{\sigma} &= 2 \begin{pmatrix}
\sigma_x^2 & \xi \\
\xi & \sigma_p^2
\end{pmatrix}, \label{eq:sdfsdf} \\
\sigma_x^2 &= \langle x^2 \rangle - \langle x \rangle^2, \label{eq:sdfsdf2222} \\
\sigma_p^2 &= \langle p^2 \rangle - \langle p \rangle^2, \label{eq:sdfsdf3333} \\
\xi &= \langle x p \rangle - \langle x \rangle \langle p \rangle - \tfrac{\mathrm{i}}{2}.  \label{eq:sdfsdf4444}
\end{align}
The determinant $\mathcal{D}$ of the covariance matrix $\boldsymbol{\sigma}$ is simply
\begin{equation}
\label{eq:ssdfsdfsdfsdfdfsdf}
\mathcal{D} = 4 \left( \sigma_x^2 \sigma_p^2 - \xi^2 \right), 
\end{equation}
which plays a key role in judging the thermodynamic performance of the bosonic quantum batteries considered in what follows. The link to Heisenberg’s uncertainty principle $\sigma_x \sigma_p \ge 1/2$, for the product of standard deviations of position $\sigma_x$ and momentum $\sigma_p$, and the given form of $\mathcal{D}$ from Eq.~\eqref{eq:ssdfsdfsdfsdfdfsdf} is most apparent.


\subsection{Measures of performance}

The canonical position and momentum operators $\hat{x}$ and $\hat{p}$, first introduced in Eq.~\eqref{eq:coxspvdsv} and Eq.~\eqref{eq:coxspvddfsfvsv}, can be decomposed into $\hat{x} = (b^\dagger + b)/\sqrt{2}$ and $\hat{p} = \mathrm{i}(b^\dagger - b)/\sqrt{2}$. Here the bosonic creation and annihilation operators are $b^\dagger = (\hat{x} - \mathrm{i} \hat{p})/\sqrt{2}$ and $b = (\hat{x} + \mathrm{i} \hat{p})/\sqrt{2}$ respectively, while the bosonic commutation relation $[b , b^\dagger] = 1$ is strictly observed. Then the Hamiltonian operator $\hat{H}_b $ of a (single-mode) continuous variable quantum battery can be simply written as
\begin{equation}
\label{eq:ssdfsdfdfsdf}
\hat{H}_b = \omega_b b^\dagger b, 
\end{equation}
where $\omega_b$ is the consistent energy level spacing, found upon climbing the infinite number of rungs of the bosonic energy ladder. Such a ubiquitous Hamiltonian as Eq.~\eqref{eq:ssdfsdfdfsdf} arises in the description of many particle and quasiparticle excitations (from photons to magnons to plasmons to polaritons). Indeed, this Hamiltonian operator has already shown great utility in explaining the experimental behaviour of the photonic quantum battery reported in Ref.~\cite{Qu2023}.

In the Fock basis, the creation operator $b^\dagger$ acts so that $b^\dagger \ket{n} = \sqrt{n+1} \ket{n+1}$, meanwhile the annihilation operator $b$ enforces the result $b \ket{n} = \sqrt{n} \ket{n-1}$, where the number state $\ket{n}$ accounts for $n = 0, 1, 2 \ldots$ excitations in the bosonic mode. In terms of these ladder operators $b^\dagger$ and $b$, the quadrature variances $\sigma_x^2$ and $\sigma_p^2$, the coherence $\xi$ and the covariance matrix determinant $\mathcal{D}$ [cf. Eq.~\eqref{eq:sdfsdf2222},~\eqref{eq:sdfsdf3333},~\eqref{eq:sdfsdf4444},~\eqref{eq:ssdfsdfsdfsdfdfsdf}] explicitly then become
\begin{align}
2\sigma_x^2 &= 1 + 2 \left(  \langle b^\dagger b \rangle -  \langle b^\dagger \rangle \langle b \rangle \right)  + \langle {b^\dagger}^2  \rangle - \langle b^\dagger \rangle^2 + \langle b^2 \rangle - \langle b \rangle^2,  \label{eq:fgfdgfdg} \\
2\sigma_p^2 &= 1 + 2 \left(  \langle b^\dagger b \rangle -  \langle b^\dagger \rangle \langle b \rangle \right)  + \langle b^\dagger \rangle^2 - \langle {b^\dagger}^2 \rangle  + \langle b \rangle^2 - \langle b^2 \rangle, \label{eq:fgfdgsdfsfsdffvvfdg} \\
2\xi &= \mathrm{i} \left( \langle {b^\dagger}^2 \rangle - \langle b^\dagger \rangle^2  + \langle b \rangle^2 -  \langle b^2 \rangle \right), \label{eq:fgfdgsdfffsfsdffvvfdg} \\
\mathcal{D} &= \left( 1 + 2  \langle b^\dagger b \rangle - 2 \langle b^\dagger \rangle \langle b \rangle \right)^2 - 4 | \langle b^2 \rangle - \langle b \rangle^2 |^2. \label{eq:fffgfdgsdfsvvfsdffvvfdg}
\end{align}
These decompositions allows for more convenient calculations to be carried out later on, when we consider specific charging protocols along with the foundational quantum battery Hamiltonian operator of Eq.~\eqref{eq:ssdfsdfdfsdf}.

The total energy $E$ stored in the quantum battery should be directly proportional to the average population of the quantum harmonic oscillator mode $b$, which can reach a maximum value $E \left( t_E \right)$ at some specific charging time $t_E$, which itself lies within the overall charging period $0 \le t \le T$. The charging power $P$, the amount of stored energy per unit time, and the maximum value of this power complete a list of four common measures of interest as follows~\cite{Campaioli2023, Downing2024b}
\begin{align}
E &= \omega_b \langle b^\dagger b \rangle, \label{eq:sfdsdfsf} \\
E \left( t_E \right) &= \max_t{\{ E(t) \}},  \label{eq:sfdsdfsf22} \\
P &= E/t, \label{eq:sfdsdfsf33} \\
P \left( t_P \right) &= \max_t{\{ P(t) \}}. \label{eq:sfdsdfsf44} 
\end{align}
The thermodynamic performance of the quantum battery can be gauged using the ergotropy $\mathcal{E}$, which measures the maximum energy which can be extracted from a quantum system in a cyclic unitary process such that it is available to do work, via the relation~\cite{Allahverdyan2004, Alicki2013}
\begin{equation}
\mathcal{E} = E - E_\beta. \label{eq:sfdsdfsf4555} 
\end{equation}
The subtracted off energy $E_\beta$ is the energy of the passive quantum state $\rho_\beta$, which is defined as the quantum state from which no energy can be extracted~\cite{Lenard1978, Pusz1978}. In general, it is difficult to obtain a simple analytic expression for this important quantity after some choice of total Hamiltonian operator.

The ergotropy $\mathcal{E}$ of a bosonic quantum battery can be readily calculated in certain situations however. If it is Gaussian, the quantum state $\rho$ associated with the quantum battery can be completely described by its vector $\boldsymbol{r}$ of first moments [cf. Eq.~\eqref{eq:coxspvdsv}] and its covariance matrix $\boldsymbol{\sigma}$ of second moments [cf. Eq.~\eqref{eq:coxspvddfsfvsv}] alone~\cite{Ferraro2005, Olivares2012}. In this interesting Gaussian case, the passive state energy $E_\beta$ can then be exactly found from the rather neat expression~\cite{Farina2019, Downing2023}
\begin{equation}
\label{eq:sfdssdfsdfsddfsf}
E_\beta = \omega_b \left( \tfrac{\sqrt{\mathcal{D}}-1}{2} \right),
\end{equation}
where the covariance matrix determinant $\mathcal{D}$ is defined in Eq.~\eqref{eq:fffgfdgsdfsvvfsdffvvfdg}, which therefore gives immediate access to the desired ergotropy $\mathcal{E}$ [cf. Eq.~\eqref{eq:sfdsdfsf4555}]. Furthermore, the purity $\mathcal{P}$ of a single-mode Gaussian state is only dependent on the quantity $\mathcal{D}$ via the simple expression $\mathcal{P} = 1/\sqrt{\mathcal{D}}$, which can take on values satisfying the inequality $0 \le \mathcal{P} \le 1$~\cite{Ferraro2005, Olivares2012}. Hence a pure Gaussian state where $\mathcal{P} = 1$ is associated with $\mathcal{D} = 1$, while mixed Gaussian states are characterized by the inequality $\mathcal{D} > 1$.

Notably, for a minimum uncertainty state satisfying Heisenberg's criterion $\sigma_x \sigma_p = 1/2$, the covariance matrix determinant collapses into $\mathcal{D} = 1$ [cf. Eq.~\eqref{eq:ssdfsdfsdfsdfdfsdf}]. The coherence should disappear ($\xi = 0$) due to the property $\mathcal{D} \ge 1$, as is guaranteed by a consideration of the purity $\mathcal{P}$. Therefore, for minimum uncertainty states the passive state energy $E_\beta$ vanishes [cf. Eq.~\eqref{eq:sfdssdfsdfsddfsf}], which results in all of the energy stored in that quantum battery being ergotropic ($E = \mathcal{E}$). This is a hallmark of a thermodynamically perfect quantum battery. As already previewed in Fig.~\ref{squeez}~(c), one can imagine that the craved for minimum uncertainty condition $\sigma_x \sigma_p = 1/2$ can be achieved through a more exotic pathway with a squeezed quadrature variance $\sigma_x^2 < 1/2$ (pink line) and $\sigma_p^2 > 1/2$ (orange line), as well as through a more standard route where $\sigma_x^2 = \sigma_p^2 = 1/2$ identically (dashed green line).


\subsection{Two-component quantum battery model}

We shall consider a class of so-called two-component quantum battery models~\cite{Pellegrini2018, Keck2019, Downing2024, Shastri2025}, defined by the overall system Hamiltonian $\hat{H}$ operator
\begin{equation}
\label{eq:sfdcdsfvcsdf}
\hat{H} = \hat{H}_a + \hat{H}_b + \hat{H}_{a-b} + \hat{H}_{d},
\end{equation}
where the bosonic battery Hamiltonian $\hat{H}_b$ was already defined in Eq.~\eqref{eq:ssdfsdfdfsdf}. We suppose that there is a charger subsystem described by the Hamiltonian $\hat{H}_a$, which interacts with the energy storing battery mode $b$ via the coupling Hamiltonian $\hat{H}_{a-b}$ only during the charging phase (which starts at the time $t = 0$ and ends at some chosen instant where $t = T$). The driving Hamiltonian $\hat{H}_d$ supplies energy to the charger, and the initial state of the system at $t = 0$ is taken to be the vacuum state without any excitations in either mode.

We imagine that the charger subsystem (mode $a$) interacts rather easily with environment such that it can be readily driven, however this utility would come at the cost of the charger suffering from dissipation. However we assume that the battery subsystem (mode $b$) is rather inaccessible, being well isolated from the environment such that all of the stored energy remains in the battery during the storage phase (that is, for times $t > T$). In this scenario, the following quantum master equation (in standard Gorini–Kossakowski–Sudarshan–Lindblad form) describes the open quantum system version of the model via~\cite{Breuer2002}
\begin{equation}
\label{eq:sjhgfdfsdf}
\partial_t \varrho = \mathrm{i} [ \varrho, \hat{H} ] + \tfrac{\gamma}{2} \left( 2 a \varrho a^\dagger - a^\dagger a \varrho - \varrho a^\dagger a \right),
\end{equation}
where $\varrho$ is the quantum state of the combined system, with the vacuum state initial condition $\varrho (t = 0) = \ket{0}_a \bra{0} _a \otimes  \ket{0}_b \bra{0} _b$ occurring at the commencement of the charging period. The dissipation rate of the charger (mode $a$) subsystem $\gamma \ge 0$, while the energy holder (mode $b$) subsystem is necessarily considered dissipationless in order to serve as a realistic battery in the storage phase ($t > T$) when the charger is disconnected (so that the coupling Hamiltonian becomes $\hat{H}_{a-b} = 0$).

This two-component quantum battery model is sketched in Fig.~\ref{squeez}~(a). The cartoon depicts the charger being driven by a laser with the amplitude $\Omega$ (green region) and suffering from losses at the rate $\gamma$ (blue region), while energy is being transferred from the charger and into the battery via some -- as yet undefined -- coupling (orange region). In what follows, we shall briefly revisit the prototypical case of linear charger--battery coupling (at the rate $g$) in Sec.~\ref{eq:linear}, before delving into the captivating case of nonlinear coupling (at the rate $J$) in Sec.~\ref{eq:Nonlinear}, both of which showcase the utility of the quantum continuous variable descriptions of Eqs.~\eqref{eq:coxspvdsv}~and~\eqref{eq:coxspvddfsfvsv}.


\section{Linear coupling}
\label{eq:linear}

Let us warm-up by considering the Hamiltonian model of Eq.~\eqref{eq:sfdcdsfvcsdf}, complete with a linear interaction between the charger and battery at some coupling rate $g$ (following on from the pioneering work of Ref.~\cite{Farina2019}). Along with Eq.~\eqref{eq:ssdfsdfdfsdf} for the bosonic battery mode $b$, the Hamiltonians accounting for the charger, coupling and driving read
\begin{align}
\label{eq:dsfsdfsdf}
\hat{H}_a &= \omega_b a^\dagger a, \\
\hat{H}_{a-b} &= g \left( a^\dagger b + b^\dagger a \right), \\
\hat{H}_d &= \Omega \left( \mathrm{e}^{- \mathrm{i} \omega_d t } a^\dagger + \mathrm{e}^{ \mathrm{i} \omega_d t }  a \right), \label{eq:dfsdfvghhh}
\end{align}
where we have supposed a driving laser of amplitude $\Omega$ and driving frequency $\omega_d$ in Eq.~\eqref{eq:dfsdfvghhh}. This setup is represented in Fig.~\ref{squeez}~(a), including the laser drive (green region) and linear coupling (orange region). The dissipative charger mode is implied by the heat bath (blue region) and is captured mathematically by the quantum master equation of Eq.~\eqref{eq:sjhgfdfsdf}.

Transforming the quantum state $\varrho$ into $\tilde{\varrho}$ with the frame change $\tilde{\varrho} = U \varrho U^\dagger$, carried out with the aid of the unitary transformation $U = \exp ( \mathrm{i} ( \hat{H}_a + \hat{H}_b ) t)$, leads to the rotated quantum master equation for $\tilde{\varrho}$ [cf. Eq.~\eqref{eq:sjhgfdfsdf}]
\begin{align}
\label{eq:sjhgfsdfsfddfsdf}
\partial_t \tilde{\varrho} &= \mathrm{i} [ \tilde{\varrho}, \hat{\mathcal{H}} ] + \tfrac{\gamma}{2} \left( 2 a \tilde{\varrho} a^\dagger - a^\dagger a \tilde{\varrho} - \tilde{\varrho} a^\dagger a \right), \\
\hat{\mathcal{H}} &=  g \left( a^\dagger b + b^\dagger a \right) + \Omega \left( a^\dagger +  a \right),
\end{align}
where we have considered the resonant case where the driving frequency $\omega_d = \omega_b$, since then the rotated Hamiltonian operator $\hat{\mathcal{H}} = U ( \hat{H}_{a-b} + \hat{H}_{d} ) U^\dagger$ simplifies. Using the trace relation $\langle \hat{\mathcal{O}} \rangle = \Tr{(\tilde{\varrho} \hat{\mathcal{O}})}$, for some arbitrary operator $\hat{\mathcal{O}}$, with the master equation of Eq.~\eqref{eq:sjhgfsdfsfddfsdf} allows one can to find the equation of motion for the mean occupation of the battery $\langle b^\dagger b \rangle$, and indeed any other relevant averaged operator, as is shown in detail in the Supplemental Material~\cite{Supplemental}.

\begin{figure*}[tb]
 \includegraphics[width=1.0\linewidth]{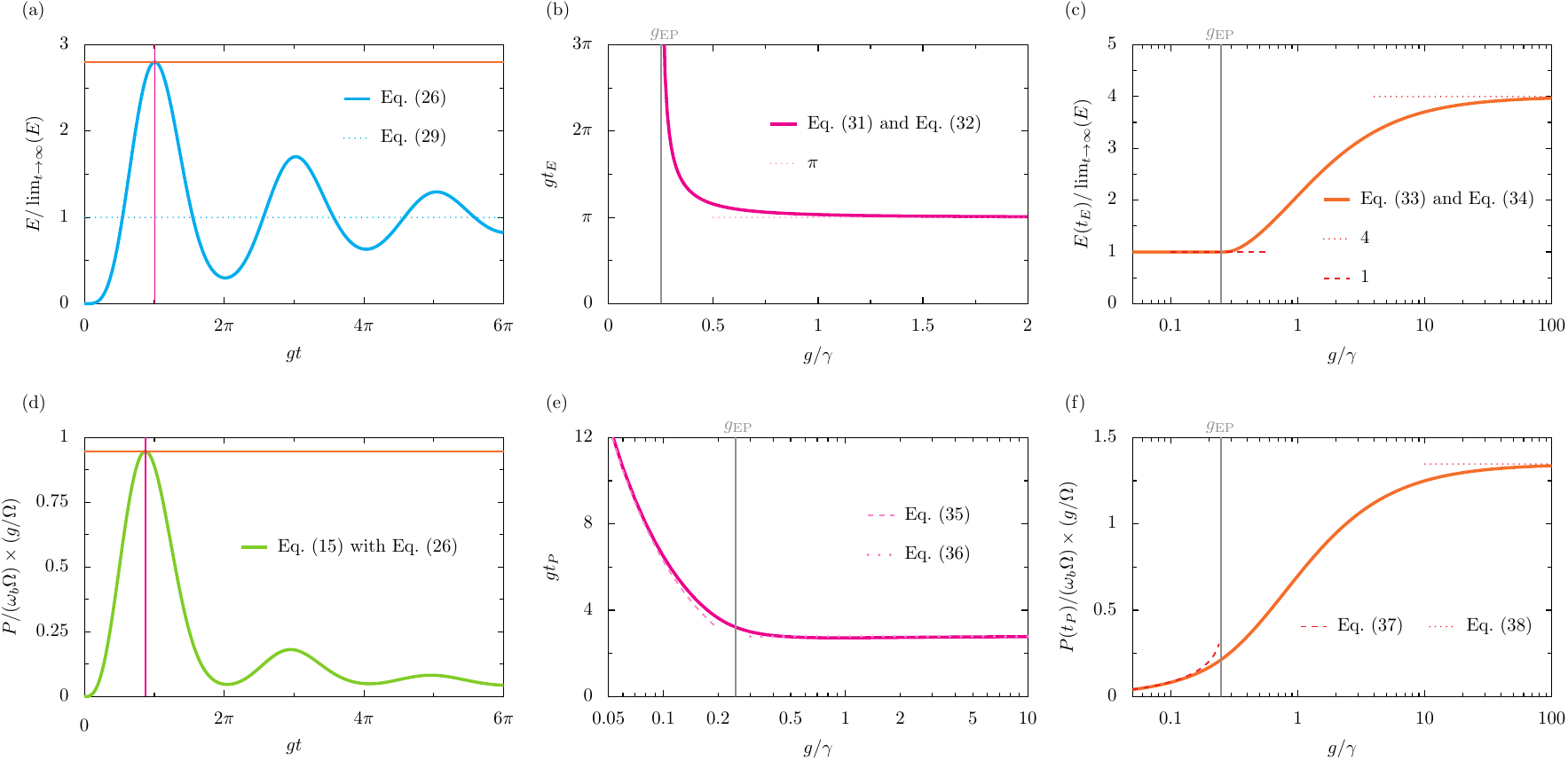}
 \caption{\textbf{Performance of the quantum battery with linear coupling.} Panel (a): the energy $E$ (cyan line) stored in the quantum battery, in units of the asymptotic result $\lim_{t \to \infty} (E)$, as a function of time $t$ (in units of $1/g$) [cf. Eq.~\eqref{eq:sdfssfddfgdfgfsdfsdf}]. Dotted line: the steady state energy [cf. Eq.~\eqref{eq:dfgdfgdg2}]. The instant in time (pink line) and specific energy (orange line) of the energetic maximum is marked. In this column we take the charger--battery coupling strength $g = \gamma/2$, where $\gamma$ is the decay rate of the charger. Panel (b): the optimal charging time $t_E$ (pink line) in units of $1/g$, as a function of the dimensionless ratio $g/\gamma$ [cf. Eq.~\eqref{eq:sdfsdfsdfsdf} and Eq.~\eqref{eq:sdfsdfsdfsdfcccc}]. Dotted line: the asymptote $t_E = \pi/g$ of the dissipationless limit. Panel (c): the maximum energy $E(t_E)$ (orange line) as a function of $g/\gamma$ [cf. Eq.~\eqref{eq:jhfghfgfgh} and Eq.~\eqref{eq:jhfghfgfgh2}]. Dashed line: the steady state result [cf. Eq.~\eqref{eq:dfgdfgdg2}]. Dotted line: in the dissipationless limit, the result is four times the steady state result [cf. Eq.~\eqref{eq:dfgdfgdg} when $t = \pi/g$]. Panel (d): the charging power $P$ (green line) of the quantum battery, in units of $\omega_b \Omega^2/g$, as a function of time $t$ [cf. Eq.~\eqref{eq:sfdsdfsf33} with Eq.~\eqref{eq:sdfssfddfgdfgfsdfsdf}]. The instant in time (pink line) and specific power (orange line) of the power maximum is marked. Panel (e): the optimal time $t_P$ (pink line) as a function of $g/\gamma$. Dashed line: the weak coupling limit [cf. Eq.~\eqref{eq:sfeessefesfsdfsdf}]. Dotted line: the strong coupling limit [cf. Eq.~\eqref{eq:sfeessefesfsdfsdf111}]. Panel (f): the maximum power $P(t_P)$ (orange line) as a function of $g/\gamma$. Dashed line: the weak coupling limit [cf. Eq.~\eqref{eq:sdfsdfssefdfsdf}]. Dotted line: the strong coupling limit [cf. Eq.~\eqref{eq:fdsvsfffffb}]. In this figure the vertical grey lines in panels (b, c, e, f) mark the location of the exceptional point $g_{\mathrm{EP}}$ [cf. Eq.~\eqref{eq:sdfsf}], and the energies $E$ referred to in the top row of panels are exactly equivalent to ergotropies $\mathcal{E}$ since the quantum battery is in a minimum uncertainty state [cf. Eq.~\eqref{eq:sdfdffhghfssfdfsdfsdf}]. }
 \label{linny}
\end{figure*}


\subsection{Energetics}

The dynamical energy $E$ stored in the linear quantum battery follows directly from Eq.~\eqref{eq:sfdsdfsf} and the mean population $\langle b^\dagger b \rangle$ of the battery, as explicity calculated in the Supplemental Material~\cite{Supplemental}. The exact analytical form for $E$ exhibits damped oscillations (with a nonzero steady state) via
\begin{equation}
\label{eq:sdfssfddfgdfgfsdfsdf}
 E = \omega_b \left( \tfrac{\Omega}{g} \right)^2 \Bigl\{ 1 - \left[ \cos \left( G t \right) + \tfrac{\gamma}{4G} \sin \left( G t \right) \right] \mathrm{e}^{- \frac{\gamma t}{4}} \Bigl\}^2, \\
\end{equation}
where there is a clear competition between the charger--battery coupling rate $g$ and the damping rate $\gamma$. Therefore, it is convenient to introduce the renormalized frequency parameter $G$, where
\begin{equation}
\label{eq:sdfsssfddfssdffddfgdfgfsdfsdf}
 G = \sqrt{g^2 - \left(\tfrac{\gamma}{4}\right)^2 }.
 \end{equation}
 An interesting borderline situation arises when the coupling strength $g = g_{\mathrm{EP}}$, a location in parameter space which corresponds to an exceptional point in the governing non-Hermitian effective Hamiltonian due to the coalescence of two eigenvalues and eigenvectors exactly at that point~\cite{Fox2023, Saroka2025}, explicitly 
 \begin{equation}
\label{eq:sdfsf}
g_{\mathrm{EP}} = \tfrac{\gamma}{4}.
 \end{equation}
 The dynamical energy $E$ stored in the quantum battery is plotted as the thick cyan line in Fig.~\ref{linny}~(a). The curve displays characteristic population cycles due to the strong charger--battery coupling, which are eventually damped out into a steady state (horizontal dotted line) as the driving and dissipation settle down. Clearly, in this strong coupling regime the nonzero dissipation present in the system means that the first energetic maxima $E(t_E)$ is also the global maximum (horizontal orange line), which suggests there is a obvious optimal charging time $t_E$ (vertical pink line) at which the charger should be disconnected from the battery mode.
 
This driven-dissipative system eventually reaches its steady state at a long enough timescale such that $t \gg 4/\gamma$, thanks to the interplay between the driving laser and lossy charger mode, whereby the steady state stored energy arises as
\begin{equation}
 \lim_{t \to \infty}{  E } = \omega_b \left( \tfrac{\Omega}{g} \right)^2. \label{eq:dfgdfgdg2} 
\end{equation}
It is apparent that this steady state energy result is not optimal for maximizing the stored energy. For example, in the dissipationless limit (where the damping rate $\gamma \to 0$) the full and dynamical stored energy $E$ of Eq.~\eqref{eq:sdfssfddfgdfgfsdfsdf} reduces to
\begin{equation}
  \lim_{\gamma \to 0}{ E } = \omega_b \left( \tfrac{2\Omega}{g} \right)^2 \sin^4 \left( \tfrac{g t}{2} \right), \label{eq:dfgdfgdg} 
\end{equation}
which displays a fourfold increase in stored energy compared to the steady state, since the energy maximum $4 \omega_b (\Omega/g)^2$ is attainable at the specific charging time $t_E = \pi/g$. In general, the optimal charging time $t_E$ [cf. Eq.~\eqref{eq:sfdsdfsf22}] depends on the coupling regime, as defined in reference to the exceptional point $g_{\mathrm{EP}}$ [cf. Eq.~\eqref{eq:sdfsf}], like so
\begin{align}
t_E  &= \infty,  &&g \le g_{\mathrm{EP}}, \label{eq:sdfsdfsdfsdf}  \\
t_E  &= \tfrac{\pi}{G}, &&g > g_{\mathrm{EP}}. \label{eq:sdfsdfsdfsdfcccc}
\end{align}
Notably, in the weak coupling regime ($g \le g_{\mathrm{EP}}$) the energetic maximum is only achievable asymptotically, while in the strong coupling regime the first energetic maxima occurs relatively quickly, scaling inversely with the renormalized coupling strength $G$ [cf. Eq.~\eqref{eq:sdfsssfddfssdffddfgdfgfsdfsdf}]. This behaviour of the optimal charging time $t_E$ is plotted as the thick pink line in Fig.~\ref{linny}~(b), with the exceptional point (vertical grey line) demarcating the two distinct regimes of response. The asymptotic result $\lim_{g \gg \gamma}{(t_E)} = \pi/g$  recovers the dissipationless result (dotted pink line) coming from Eq.~\eqref{eq:dfgdfgdg}.

The two cases of optimal charging time $t_E$, given explicitly in Eq.~\eqref{eq:sdfsdfsdfsdf} and Eq.~\eqref{eq:sdfsdfsdfsdfcccc}, are associated with their respective optimal stored energies $E ( t_E )$ as follows [cf. Eq.~\eqref{eq:sfdsdfsf22}] 
\begin{align}
E \left( t_E \right) &=  \omega_b \left( \tfrac{\Omega}{g} \right)^2,  \quad\quad &&g \le g_{\mathrm{EP}}, \label{eq:jhfghfgfgh} \\
E \left( t_E \right) &=  \omega_b \left( \tfrac{\Omega}{g} \right)^2 \left( 1 + \mathrm{e}^{-\frac{\pi \gamma}{4G}} \right)^2, &&g > g_{\mathrm{EP}}. \label{eq:jhfghfgfgh2}
\end{align}
These optimal energy relations are plotted in Fig.~\ref{linny}~(c) as the thick orange line, alongside the limiting cases of very strong coupling (dotted line) and very weak coupling (dashed line). This curve showcases how the fourfold increase in stored energy (compared to the steady state) is approached with increasing coupling strength $g$. Notably, in the very strong coupling regime where $g \gg g_{\mathrm{EP}}$, this ideal quadrupling in stored energy is slowly degraded like $E \left( t_E \right) \simeq 4 \omega_b (\Omega/g)^2 (1 - \pi \gamma / 4 G)$. This brief energetic analysis highlights the importance of achieving strong coupling for both peak energetic and temporal performance, as well as the importance of correctly judging the charging time $t_E$ at which one should ideally disconnect the charger mode from the battery mode.


\subsection{Charging power}

The associated charging power $P$ follows immediately from Eq.~\eqref{eq:sfdsdfsf33} with Eq.~\eqref{eq:sdfssfddfgdfgfsdfsdf}, which judges the combined energetic and temporal performance of the quantum battery in one simple measure. A typical result in the strong coupling regime is plotted as the solid green line in Fig.~\ref{linny}~(d). This curve showcases a well-defined peak in power $P(t_P)$ (horizontal orange line) at a specific charging time $t_P$ (vertical pink line). Eventually, the dissipation present in the coupled system -- as well as the always advancing charging time -- wash out the amount of energy stored per unit time to essentially zero.

The optimal charging time $t_P$ [cf. Eq.~\eqref{eq:sfdsdfsf44}] can be readily found by analyzing the turning points of the charging power $P$, and the result is plotted as the thick pink line in Fig.~\ref{linny}~(e). This temporal behaviour, which is strikingly different below and above the exceptional point $g_{\mathrm{EP}}$ (vertical gray line), can be captured analytically in the limiting cases of very weak and very strong coupling. When $g \ll \gamma$, the optimal charging time $t_P$ acts like
\begin{equation}
\label{eq:sfeessefesfsdfsdf}
t_P  = 
A \frac{\gamma}{2 g^2}, \quad\quad\quad g \ll \gamma, 
\end{equation}
where the dimensionless number $A = 1.256 \ldots$ arises from the expression $A = -1/2 - W_{-1}(\alpha)$, which itself comes from a turning point equation. The argument of the special function $\alpha = -1/(2\sqrt{\mathrm{e}})$. Here $W_{m}(x)$ is the Lambert W-function, sometimes called the product logarithm function, which is a multivalued function of argument $x$ (with infinitely many branches, which are indexed by the integer $m$). The expression of Eq.~\eqref{eq:sfeessefesfsdfsdf} is plotted as the dashed pink line in Fig.~\ref{linny}~(e), illustrating that the optimal time $t_P$ has an inverse-square divergence with small coupling $g$, since under weak coupling more charging time is needed before the peak in power is finally reached. In the opposing and desirable limit of $g \gg \gamma$, when one can completely neglect dissipation, the optimal time $t_P$ reads
\begin{equation}
\label{eq:sfeessefesfsdfsdf111}
t_P  =   \frac{B}{g}, \quad\quad\quad g \gg \gamma,
\end{equation}
where the dimensionless number $B = 2.786\ldots$ is a solution to the transcendental turning point equation $\tan(B/2) = 2B$. The result of Eq.~\eqref{eq:sfeessefesfsdfsdf111} is plotted as the dotted pink line in Fig.~\ref{linny}~(e), which shows that under strong coupling the maximum charging power is rather appealingly arrived at ever more quickly with stronger coupling $g$, following an inverse-linear relationship. Further analysis of the two charging times $t_E$ and $t_P$ is provided elsewhere~\cite{Supplemental}.
 
The maximum charging power $P ( t_P )$ which can be achieved in this linear quantum battery, for some dimensionless ratio of the coupling to dissipation rate $g/\gamma$, is given as the thick orange line in Fig.~\ref{linny}~(f). This performative curve can be understood using the asymptotic optimal charging times $t_P$ from Eq.~\eqref{eq:sfeessefesfsdfsdf} and Eq.~\eqref{eq:sfeessefesfsdfsdf111} as follows
\begin{align}
P \left( t_P \right) &= C \omega_b \frac{\Omega^2}{\gamma},  &&g \ll \gamma, \label{eq:sdfsdfssefdfsdf}  \\
P \left( t_P \right) &=  D \omega_b \frac{\Omega^2}{g}, &&g \gg \gamma, \label{eq:fdsvsfffffb}
\end{align}
where the numerical prefactors $C = 2 [ 1 - \exp(-A) ]^2/A = 0.814 \dots$ and $D = 4 \sin^4(B/2)/B = 1.347 \ldots$ arise from the aforementioned limiting forms of the charging times $t_P$. With very weak coupling $g \ll \gamma$, Eq.~\eqref{eq:sdfsdfssefdfsdf} encapsulates the lower bound on the maximum power, while for very strong coupling $g \gg \gamma$, Eq.~\eqref{eq:fdsvsfffffb} estimates the kind of charging powers which can likely be achieved within this simple linear model.


\subsection{Thermodynamics}

The second moments of the bosonic quantum battery with linear coupling factorize exactly, for example $\langle b^\dagger b \rangle = \langle b^\dagger \rangle \langle b \rangle$, as is shown in the Supplemental Material~\cite{Supplemental}. This immediately means that determinant $\mathcal{D}$ of the covariance matrix [cf. Eq.~\eqref{eq:fffgfdgsdfsvvfsdffvvfdg}] evaluates to unity, that is
\begin{equation}
\label{eq:dfgsdsssddfgdfgfdfdfsdfddsvvdfg}
\mathcal{D} = 1.
\end{equation}
Along with the expression of Eq.~\eqref{eq:sfdssdfsdfsddfsf}, which holds here due to the Gaussian nature of the linear battery, the elegant result of Eq.~\eqref{eq:dfgsdsssddfgdfgfdfdfsdfddsvvdfg} ensures that the passive state energy $E_\beta$ vanishes identically ($E_\beta = 0$). Then all of the energy $E$ stored in the quantum battery is ergotropic [cf. Eq.~\eqref{eq:sfdsdfsf4555}] or
\begin{equation}
\label{eq:sdfdffhghfssfdfsdfsdf}
 E = \mathcal{E}.
\end{equation}
Thermodynamically, this Gaussian quantum battery is then (in a sense) perfectly efficient, essentially due to the form of the influential determinant of Eq.~\eqref{eq:dfgsdsssddfgdfgfdfdfsdfddsvvdfg}. Using the decomposition of Eq.~\eqref{eq:ssdfsdfsdfsdfdfsdf}, this mathematical result can also be traced back to the the quadrature variances $\sigma_x^2$ and $\sigma_p^2$ being pinned at the values [cf. Eq.~\eqref{eq:fgfdgfdg} and Eq.~\eqref{eq:fgfdgsdfsfsdffvvfdg}]
\begin{equation}
\label{eq:sdfshfhdddf}
\sigma_x^2 = \sigma_p^2 = \tfrac{1}{2}, 
\end{equation}
as well as the coherence vanishing or $\xi = 0$ [cf. Eq.~\eqref{eq:fgfdgsdfffsfsdffvvfdg}]. In this way, the fact that the total stored energy $E$ is completely ergotropic stems from the bosonic quantum battery with linear coupling remaining in a minimum uncertainty quantum state, that where the relation $\sigma_x \sigma_p = 1/2$ holds, throughout the duration of the charging phase from the initial vacuum state preparation at $t = 0$ to the disconnection of the charger at $t = T$. This proof of the ergotropic efficiency of the linear quantum battery concludes our groundwork with the linear coupling case, and lays the foundations for the physically richer nonlinear case we consider next.


\section{Nonlinear coupling}
\label{eq:Nonlinear}

Let us now consider an arguably more interesting problem, that of nonlinear coupling between the charger and battery at some coupling rate $J$. The full Hamiltonian model of Eq.~\eqref{eq:sfdcdsfvcsdf} [along with the battery mode $b$ of Eq.~\eqref{eq:ssdfsdfdfsdf}] is made explicit with the following charger, coupling and driving Hamiltonians 
\begin{align}
\hat{H}_a &= 2\omega_b a^\dagger a, \label{eq:dsfssdfsdfdfsdf} \\
\hat{H}_{a-b} &= J \left( a^\dagger  b b + b^\dagger b^\dagger a \right), \label{eq:dsfssccccc} \\
\hat{H}_d &= \Omega \left( \mathrm{e}^{- 2\mathrm{i} \omega_d t } a^\dagger + \mathrm{e}^{2 \mathrm{i} \omega_d t }  a \right).
\end{align}
Importantly, compared to the linear model of Sec.~\ref{eq:linear}, here the energy level spacing in the charger mode is doubled to $2\omega_b$, and  (while the laser remains associated with the amplitude $\Omega$) the driving laser frequency is also doubled to $2\omega_d$. This nonlinear quantum battery arrangement is sketched in Fig.~\ref{squeez}~(a), complete with the laser drive (green region), nonlinear interaction (orange region) and dissipation from the charger mode $a$ into a heat bath (blue region), which is again modelled with the quantum master equation of Eq.~\eqref{eq:sjhgfdfsdf}.

The undriven version of the decidedly nonquadratic Hamiltonian operator of Eq.~\eqref{eq:sfdcdsfvcsdf} is $\hat{H}_u$, where $\hat{H}_u = \hat{H}_a + \hat{H}_b + \hat{H}_{a-b}$, which does not conserve the excitation number due to the explicitly nonlinear coupling featuring in Eq.~\eqref{eq:dsfssccccc}. This fact can be understood from the commutator result $[ \hat{N}, \hat{H}_u  ] \ne 0$, where the total number operator $\hat{N} = a^\dagger a + b^\dagger b$. However, the related object $\hat{M} = 2 a^\dagger a + b^\dagger b$ is indeed conserved since this commutator vanishes, or $[ \hat{M}, \hat{H}_u  ] = 0$. This exact cancellation corresponds to every emitted photon of frequency $2\omega_b$ (coming from the charger oscillator, mode $a$) leading to the absorption of two photons of equal frequency $\omega_b$ (in the battery oscillator, mode $b$). The cartoon of Fig.~\ref{squeez}~(b) suggests such a nonlinear coupling process, which is reminiscent of degenerate parametric downconversion (mediated perhaps by an appropriately nonlinear crystal) as explored for example in the experiments reported in Refs.~\cite{Guo2017, Sandbo2020}.

Working with a rotated density matrix $\tilde{\varrho} = U \varrho U^\dagger$, where the unitary transformation $U = \exp ( \mathrm{i} ( \hat{H}_a + \hat{H}_b ) t)$, the quantum master equation of Eq.~\eqref{eq:sjhgfdfsdf} collapses into
\begin{align}
\label{eq:sjhgfsdasdasdfsfddfsdf}
\partial_t \tilde{\varrho} &= \mathrm{i} [ \tilde{\varrho}, \hat{\mathcal{H}} ] + \tfrac{\gamma}{2} \left( 2 a \tilde{\varrho} a^\dagger - a^\dagger a \tilde{\varrho} - \tilde{\varrho} a^\dagger a \right), \\
\hat{\mathcal{H}} &=  J \left( a^\dagger b b + b^\dagger  b^\dagger a \right) + \Omega \left( a^\dagger +  a \right),
\end{align}
where we have considered the resonant case where the driving frequency $\omega_d = \omega_b$, since then the rotated Hamiltonian operator $\hat{\mathcal{H}} = U ( \hat{H}_{a-b} + \hat{H}_{d} ) U^\dagger$ simplifies. As demonstrated in the Supplemental Material~\cite{Supplemental}, the equations of motion for the statistical moments of the quantum battery system can then be found, giving access to the average population of the battery $\langle b^\dagger b \rangle$ for example. It is also shown there that the first moments $\langle b \rangle$ and $\langle b^\dagger \rangle$ are necessarily zero, so that the covariance matrix determinant $\mathcal{D}$ of Eq.~\eqref{eq:fffgfdgsdfsvvfsdffvvfdg} reduces to the even simpler form
\begin{equation}
\label{eq:dfgsdssdfsdfddsvvdfg}
\mathcal{D} = \left( 1 + 2  \langle b^\dagger b \rangle  \right)^2 - 4 | \langle b b \rangle  |^2.
\end{equation}
Furthermore, the first derivative (with respect to time) of this influential object $\mathcal{D}$, divided by a factor of four for convenience, is given by the expression
\begin{equation}
\label{eq:sdfhhtht}
\tfrac{1}{4} \partial_t \mathcal{D} = \left( 1 +2 \langle b^\dagger b \rangle \right) \partial_t \langle b^\dagger b \rangle
- \langle b^\dagger b^\dagger \rangle \partial_t \langle b b \rangle 
- \langle b b \rangle \partial_t \langle b^\dagger b^\dagger \rangle.
\end{equation}
This derivative is useful in analysing the thermodynamic behaviour of the nonlinear quantum battery in what follows.


\subsection{Cumulant expansion}

The nonlinear nature of this problem ensures that the equations of motion of the moments form an infinite hierarchy, with lower-order moments always depending upon higher-order moments in perpetuity. This somewhat unwieldy system of equations can plausibly be truncated by performing a cumulant expansion~\cite{Kubo1962, Kubo1963}, which (at the second-order of approximation) gives rise to just five governing equations 
\begin{align}
\label{eq:sddfsfhhfsdf}
 \partial_t \langle a \rangle &= - \tfrac{\gamma}{2}  \langle a \rangle - \mathrm{i}  J  \langle b b \rangle - \mathrm{i}  \Omega, \\
\partial_t \langle a^\dagger a \rangle &= - \gamma \langle a^\dagger a \rangle + 2 J \operatorname{Im} \left(  \langle a^\dagger \rangle \langle b b \rangle \right) - 2 \Omega \operatorname{Im}  \langle a \rangle, \\
 \partial_t \langle b^\dagger b \rangle &= - 4 J \operatorname{Im} \left( \langle a^\dagger \rangle \langle b b \rangle \right), \\
\partial_t \langle a a \rangle &= - \gamma \langle a a \rangle - 2 \mathrm{i} \Omega \langle a \rangle - 2 \mathrm{i} J \langle a \rangle \langle b b \rangle, \\
\partial_t \langle b b \rangle &= - 2 \mathrm{i} J \langle a \rangle - 4 \mathrm{i} J \langle a \rangle \langle b^\dagger b \rangle, \label{eq:sfdfvfvvggg}
\end{align}
as derived in detail in the Supplemental Material~\cite{Supplemental}. This approximation is indeed a reasonable one when the drive amplitude $\Omega$ is not too large compared to the coupling rate $J$, since this means that higher excitation sectors of the coupled system are left essentially unpopulated, such that the truncation made (and the discarding of higher-order correlations) is rather unconsequential~\cite{Kubo1962, Kubo1963}.

Substitution of these equations of motion [cf. Eq.~\eqref{eq:sddfsfhhfsdf}--Eq.~\eqref{eq:sfdfvfvvggg}] into the formula of Eq.~\eqref{eq:sdfhhtht} reveals a neat result for the time derivative of the covariance matrix determinant $\mathcal{D}$, explicitly
\begin{equation}
\label{eq:sdfhjliijhtht}
\partial_t \mathcal{D} = 0.
\end{equation}
Hence the determinant $\mathcal{D}$ itself is a constant in time over the charging period, starting from $t = 0$ and ending at $t = T$ when the charger is disconnected. Since the quantum battery is initialized in the vacuum state at $t = 0$ all of the moments appearing in Eq.~\eqref{eq:fffgfdgsdfsvvfsdffvvfdg} must be zero at that particular instant, such that the constant implied by the integration of Eq.~\eqref{eq:sdfhjliijhtht} must be unity, whence one may conclude
\begin{equation}
\label{eq:dfgsdsssdfdfdfsdfddsvvdfg}
\mathcal{D} = 1. 
\end{equation}
Within the second-order cumulant approximation, strings of three operators like $\langle A B C \rangle$ are discarded~\cite{Kubo1962, Kubo1963}. The largest surviving correlators are then of the form $\langle A B \rangle$, that is the second moments of the system. Hence the considered cumulant expansion forces a description of the nonlinear quantum battery using only first moments and second moments, in a similar manner as to how Gaussian states can be fully accounted for using Eq.~\eqref{eq:coxspvdsv} and Eq.~\eqref{eq:coxspvddfsfvsv} only. Therefore, within this specific approximation, it is reasonable to assume that the the passive state energy $E_\beta$ formula of Eq.~\eqref{eq:sfdssdfsdfsddfsf} essentially holds. The neat result of Eq.~\eqref{eq:dfgsdsssdfdfdfsdfddsvvdfg} then implies that the passive state energy $E_\beta = 0$, such that the stored energy $E$ is thermodynamically perfect, being from Eq.~\eqref{eq:sfdsdfsf4555}
\begin{equation}
\label{eq:dfgdg}
 E = \mathcal{E},
\end{equation}
at least within this second-order cumulant approximation of course. This useful discovery provides a nontrivial example of ergotropy optimization in a nonlinear system, complementing the elementary linear model consideration of Sec.~\ref{eq:linear}.


\subsection{Steady state}

The steady state of the system, whereby the left-hand sides of Eqs.~\eqref{eq:sddfsfhhfsdf}--\eqref{eq:sfdfvfvvggg} are necessarily zero due to the moments being unchanged in time, reveals that the long-time behaviour of the nonlinear quantum battery is described by
\begin{align}
\label{eq:dfgsdsddsvvdfg}
\lim_{t \to \infty}  \langle a \rangle &= \lim_{t \to \infty}  \langle a^\dagger a \rangle  = \lim_{t \to \infty}  \langle a a \rangle = 0, \\
\lim_{t \to \infty} \langle b b  \rangle &= -\frac{\Omega}{J}. \label{eq:dfgsdsfsdccsscddsvvdfg}
\end{align}
Then, since the relation of Eq.~\eqref{eq:dfgsdsssdfdfdfsdfddsvvdfg} holds under the second-order cumulant expansion, inverting the definitional result of Eq.~\eqref{eq:dfgsdssdfsdfddsvvdfg} and substituting in Eq.~\eqref{eq:dfgsdsfsdccsscddsvvdfg} gives access to the mean value of the battery population $\langle b^\dagger b  \rangle$. Hence the total energy $E$ stored in the nonlinear quantum battery in the steady state now follows directly from Eq.~\eqref{eq:sfdsdfsf} as
\begin{equation}
\label{eq:sssdfsfd}
\lim_{t \to \infty} E = \frac{\omega_b}{2} \left( \sqrt{1 + \left( \tfrac{2\Omega}{J} \right)^2} - 1\right),
\end{equation}
which is, quite remarkably, independent of the charger damping rate $\gamma$ -- just like the linear coupling result of Eq.~\eqref{eq:dfgdfgdg2}. The analytic expression of Eq.~\eqref{eq:sssdfsfd} immediately reveals that the lower energetic bound displays a quadratic scaling like $\lim_{t \to \infty} (E) \simeq \omega_b (\Omega / J)^2$ when $\Omega \ll J$, while the upper energetic bound instead exhibits the linear relationship $\lim_{t \to \infty} (E) \simeq \omega_b (\Omega / J)$ when $\Omega \gg J$.

Similar to the discovery of Eq.~\eqref{eq:sssdfsfd}, the quadrature variances $\sigma_x^2$ and $\sigma_p^2$ in the steady state, as defined in Eq.~\eqref{eq:fgfdgfdg} and Eq.~\eqref{eq:fgfdgsdfsfsdffvvfdg}, evaluate to the rather compact forms
\begin{align}
\lim_{t \to \infty} \sigma_x^2 &= \frac{1}{2} \left( \sqrt{ 1 + \left( \frac{2\Omega}{J} \right)^2 } - \frac{2\Omega}{J} \right), \label{eq:xfgxsdfdfg} \\
\lim_{t \to \infty} \sigma_p^2 &= \frac{1}{2} \left( \sqrt{ 1 + \left( \frac{2\Omega}{J} \right)^2 } + \frac{2\Omega}{J} \right),  \label{eq:xfgxsdfsdddfg}
\end{align}
such that the nonlinear quantum battery is seen to be in a minimum uncertainty state, since the product of standard deviations is identically $\lim_{t \to \infty} ( \sigma_x \sigma_p ) = 1/2$. Notably, unlike the linear quantum battery [cf. Eq.~\eqref{eq:sdfshfhdddf}] the minimum uncertainty in the nonlinear quantum battery is not achieved equally amongst the two quadratures, but instead the uncertainty saturation is attained through quantum squeezing~\cite{Walls1983, Loudon1987} in the $\hat{x}$-quadrature only [cf. Eq.~\eqref{eq:xfgxsdfdfg}]. This quantum squeezing phenomena is illustrated in Fig.~\ref{squeez}~(c), which shows the variances for the $\hat{x}$-quadrature (pink line) and $\hat{p}$-quadrature (orange line), in the steady state and as a function of the drive amplitude--coupling rate ratio $\Omega/J$. The minimum uncertainty relation $\sigma_x \sigma_p = 1/2$ (dashed green line) is shown as a guide for the eye. When the driving is weak and $\Omega \ll J$, the variances deviate from one half linearly like $\lim_{t \to \infty} ( \sigma_x^2 ) \simeq 1/2 -\Omega/J$ and $\lim_{t \to \infty} ( \sigma_p^2 ) \simeq 1/2 + \Omega/J$, as shown in Fig.~\ref{squeez}~(c). Conversely, in the limit where the driving is strong and so $\Omega \gg J$, the pair of variances approach the expressions $\lim_{t \to \infty} ( \sigma_x^2 ) \simeq J/8\Omega$ and $\lim_{t \to \infty} ( \sigma_p^2 ) \simeq 2 \Omega/J$, such that the $\hat{x}$-quadrature squeezing is even more apparent. This brief steady state analysis provides a concrete example of a nontrivial minimum uncertainty state leading to a thermodynamically perfect bosonic quantum battery.

\begin{figure*}[tb]
 \includegraphics[width=1.0\linewidth]{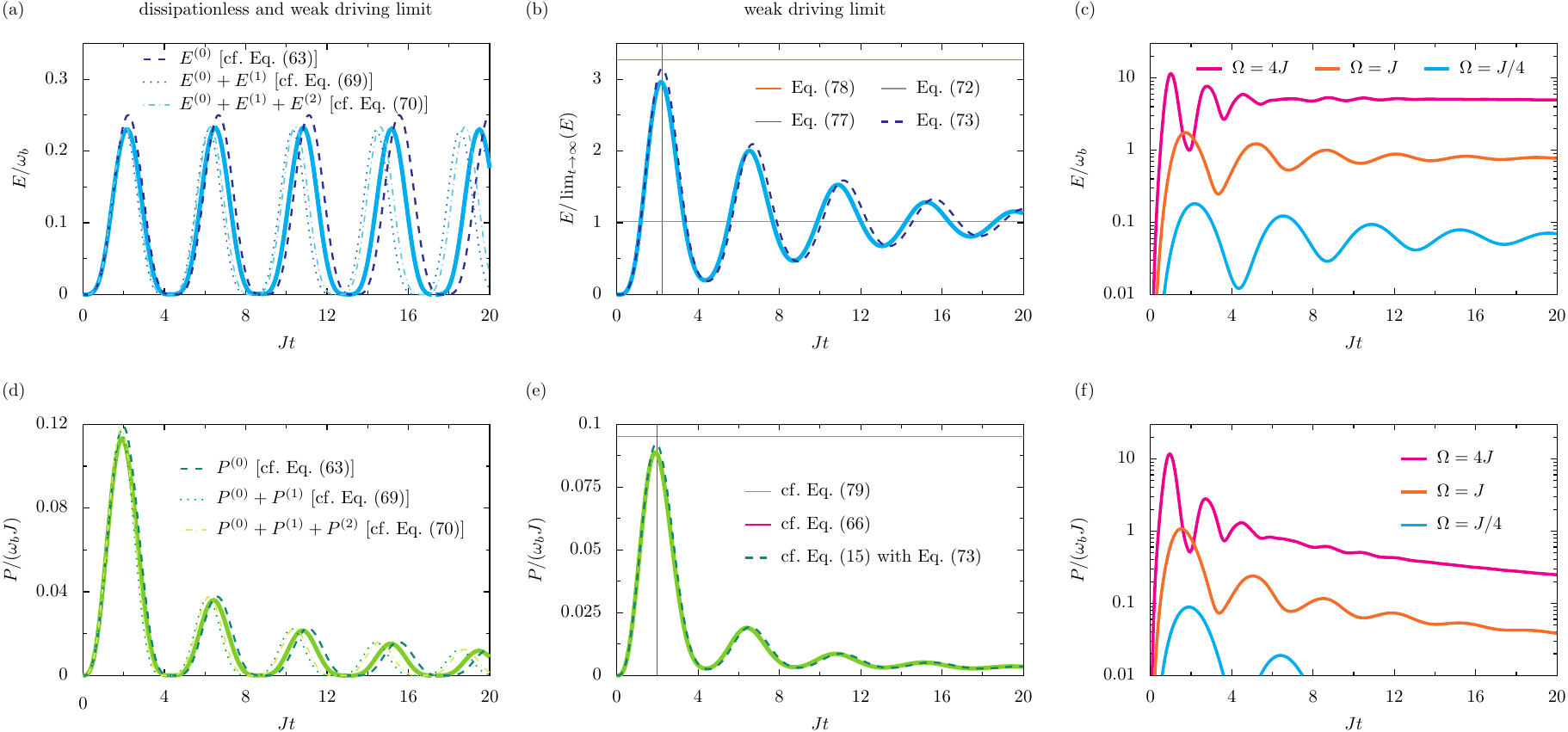}
 \caption{\textbf{Dynamics of the quantum battery with nonlinear coupling.} Panel (a): the energy $E$ stored in the quantum battery in the dissipationless limit (solid cyan line) and in units of the energy level spacing $\omega_b$, as a function of time $t$ (in units of $1/J$, the inverse of the coupling rate $J$). Dashed, dotted and dash-dotted lines: three approximate results, derived with perturbation theory of increasing order [cf. Eq.~\eqref{eq:sdfsdfs}, along with Eq.~\eqref{eq:dfgsdfdsdfgdg}, Eq.~\eqref{eq:wilf} and Eq.~\eqref{eq:sfvgvwwwq}]. Panel (b): the stored energy $E$ (solid cyan line) in units of the steady state energy, within the weak driving regime $\Omega \ll J$. Dashed blue line: the approximate behaviour [cf. Eq.~\eqref{eq:ssssdfsdfsdfdfsfd}]. Vertical pink line: the optimal charging time $t_E$ [cf. Eq.~\eqref{eq:sfdsfdfdf}]. Horizontal orange line: the approximate maximum stored energy $E(t_E)$ [cf. Eq.~\eqref{eq:sfdsfdfdffkuuuf}]. Thin grey line: the approximate steady state energy $\lim_{t \to \infty} (E)$ [cf. Eq.~\eqref{eq:sfdssdsdffsdf}]. Panel (c): a semi-logarithmic plot of the dynamic energy $E$ in units of $\omega_b$. Panels (d, e, f): the charging power $P$ of the quantum battery, in units of $\omega_b J$, corresponding to the stored energies of the above panels (a, b, c) respectively [cf. Eq.~\eqref{eq:sfdsdfsf33}]. In panel (e), the vertical pink line corresponds to the optimal charging time $t_P$ [cf. Eq.~\eqref{eq:zvxxczxczcx}] and the horizontal orange line represents the maximum power $P(t_p)$ [cf. Eq.~\eqref{eq:gfgdbdg}]. Within this figure, in the first two columns $\Omega = J/4$ and in the final two columns $\gamma = J/2$. }
 \label{nonlinny}
\end{figure*}


\subsection{Dissipationless and weak driving limit}
\label{eq:fevtnyny}

Let us now consider the dynamical behaviour of the nonlinear quantum battery in the dissipationless limit ($\gamma \to 0$), using the equations of motion given by Eqs.~\eqref{eq:sddfsfhhfsdf}--\eqref{eq:sfdfvfvvggg}. Upon considering the weak driving limit where $\Omega \ll J$, a perturbation theory can be built around the small and dimensionless parameter $\Omega/J$. Then the energy $E$ stored in the battery can be expanded in a series as
\begin{equation}
\label{eq:sdfsdfs} 
E = E^{(0)} + E^{(1)} + E^{(2)} + \cdots
\end{equation}
as discussed in detail within the Supplemental Material~\cite{Supplemental}. The leading order contribution is quadratic in the parameter $\Omega/J$, being
\begin{equation}
\label{eq:dfgsdfdsdfgdg} 
E^{(0)} = \omega_b \left( \frac{2\Omega}{J} \right)^2 \sin^4 \left( \frac{J t}{\sqrt{2}} \right), 
\end{equation}
which, similarly to the linear case of Eq.~\eqref{eq:dfgdfgdg}, displays a sine to the power four functional form as part of the flip-flop population cycles between the charger mode and the battery mode. The optimal charging time for energy $t_E$, and the maximum possible stored energy $E ( t_E )$, in this zeroth-order regime are simply [cf. Eq.~\eqref{eq:sfdsdfsf22}]
\begin{align}
\label{eq:wafcawdf}
 t_E &= \frac{\pi}{\sqrt{2} J}, \\
E \left( t_E \right) &= \omega_b \left( \frac{2\Omega}{J} \right)^2. \label{eq:wafcawdf2}
\end{align}
The energy dynamics according to Eq.~\eqref{eq:dfgsdfdsdfgdg} are plotted as the dashed dark blue line in Fig.~\ref{nonlinny}~(a), alongside the exact result (solid cyan line) for the example case where the driving amplitude $\Omega = J/4$. The optimal values of Eq.~\eqref{eq:wafcawdf} and Eq.~\eqref{eq:wafcawdf2} are indeed quite realistic approximations, as can be noticed in the chosen case of Fig.~\ref{nonlinny}~(a) where the dimensionless number estimates are $J t_E = \pi/\sqrt{2} \simeq 2.22$ and $E ( t_E ) / \omega_b = 1/4$.

Similarly, the charging power $P$ for this simple case is shown in the below panel of Fig.~\ref{nonlinny}~(d) [cf. Eq.~\eqref{eq:sfdsdfsf33}], where the exact result is given by the solid light-green line and the leading-order approximation [following Eq.~\eqref{eq:dfgsdfdsdfgdg}] by the dashed dark-green line. The optimal charging time for power $t_P$, and the maximum power $P ( t_P )$, in this zeroth-order case are given by the expressions [cf. Eq.~\eqref{eq:sfdsdfsf44}]
\begin{align}
\label{eq:zvxxczxczcx}
 t_P &= \frac{\sqrt{2} \alpha}{J}, \\
 P \left( t_P \right) &= \beta \omega_b \frac{\Omega^2}{J},  \label{eq:zvxxczxczcx2}
\end{align}
where the two dimensionless prefactors are $\sqrt{2} \alpha = 1.97\ldots$ and $\beta = 2 \sqrt{2} \sin^4 ( \alpha ) / \alpha = 1.905\ldots$ since the influential number $\alpha = 1.393\ldots$ arises as a solution to the transcendental turning point equation $\tan (\alpha) = 4 \alpha$. The analytical expressions of Eq.~\eqref{eq:zvxxczxczcx} and Eq.~\eqref{eq:zvxxczxczcx2} can be seen in Fig.~\ref{nonlinny}~(d) to be excellent approximations in this dissipationless and weak driving regime, since the zeroth order predictions are for the dimensionless numbers $J t_P = \sqrt{2}\alpha \simeq 1.97$ and $P(t_P)/\omega_b J = \beta (\Omega/J)^2 \simeq 0.119$ in this chosen case.

The leading-order perturbative result of Eq.~\eqref{eq:dfgsdfdsdfgdg} can be improved upon using the Poincaré–Lindstedt method to remove the problematic secular terms which appear later on in the asymptotic series solution~\cite{Strogatz2001}. This approach leads to the first three terms in the energetic expansion $E^{(n)}$ of Eq.~\eqref{eq:sdfsdfs} as follows
\begin{widetext}
\begin{align}
\label{eq:dfgsdfdssdsdfsfddfgdg} 
E^{(0)} &= \omega_b \left( \tfrac{2\Omega}{J} \right)^2 \sin^4 \left( \tfrac{J \tau}{\sqrt{2}} \right), \\
 E^{(1)} &= \tfrac{\omega_b}{6} \left( \tfrac{\Omega}{J} \right)^4 \sin^3 \left( \tfrac{J \tau}{\sqrt{2}} \right) \left[ 3 \sin \left( \tfrac{5 J \tau}{\sqrt{2}} \right) - 25 \sin \left( \tfrac{3 J \tau}{\sqrt{2}} \right) - 60 \sin \left( \tfrac{ J \tau}{\sqrt{2}} \right)  \right], \label{eq:wilf} \\
  E^{(2)} &= \tfrac{\omega_b}{1440} \left( \tfrac{\Omega}{J} \right)^6 \sin^4 \left( \tfrac{J \tau}{\sqrt{2}} \right) \left[ 101983 + 75156 \cos \left( \sqrt{2} J \tau \right) - 2586 \cos \left( 2\sqrt{2} J \tau \right) - 2248 \cos \left( 3\sqrt{2} J \tau \right) + 135 \cos \left( 4\sqrt{2} J \tau \right)  \right], \label{eq:sfvgvwwwq} 
\end{align}
\end{widetext}
as derived in the Supplemental Material~\cite{Supplemental}, and where (at the second-order of approximation) the shifted time variable $\tau$ is given by the polynomial expression
\begin{equation}
 \tau = \left( 1 + 5 \left( \tfrac{\Omega}{2 J} \right)^2 - \tfrac{229}{4} \left( \tfrac{\Omega}{2 J} \right)^4 \right) t.
 \end{equation} 
 These three approximate results $E^{(n)}$ are plotted in the dynamic energy $E$ and charging power $P$ plots of Fig.~\ref{nonlinny}~(a, d), where the dotted lines are for the first-order $E^{(1)}$ case and the dash-dotted lines are for second-order $E^{(2)}$ expression. These two graphs showcase the claimed improvements to the zeroth-order $E^{(0)}$ results (dashed lines) and the tendencies towards the exact results (thick solid lines).

\begin{figure*}[tb]
 \includegraphics[width=1.0\linewidth]{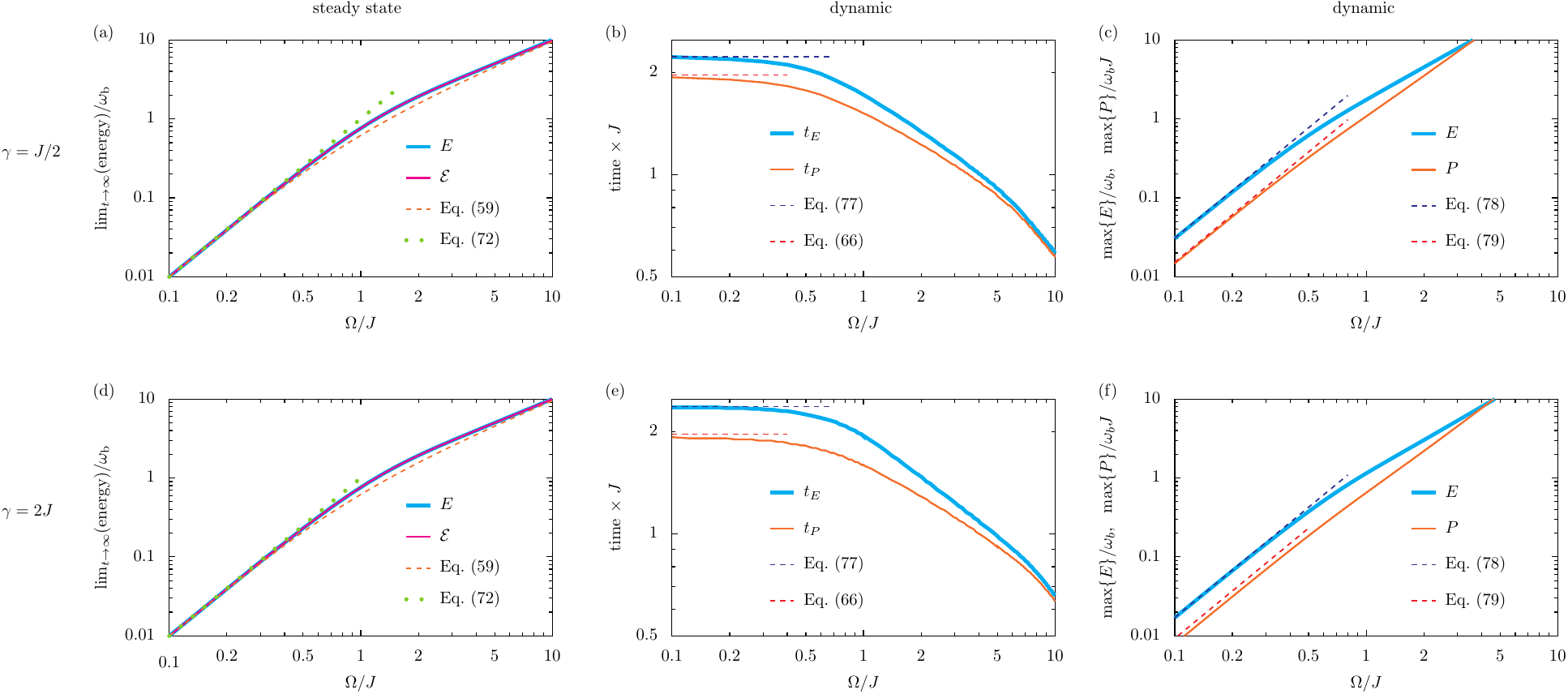}
 \caption{\textbf{Performance of the quantum battery with nonlinear coupling.} Panel (a): the energy $E$ (thick cyan line) and ergotropy $\mathcal{E}$ (thin pink line) stored in the quantum battery in the steady state, shown as a function of the drive amplitude $\Omega$ (in units of the coupling rate $J$). Dashed orange line: the cumulant approximation [cf. Eq.~\eqref{eq:sssdfsfd}]. Dotted green line: the weak driving approximation [cf. Eq.~\eqref{eq:sfdssdsdffsdf}]. Panel (b): the optimal charging times for the the quantum battery in order to maximize the energy $t_E$ (thick cyan line) and power $t_P$ (thin orange line) [cf. Eq.~\eqref{eq:sfdsdfsf22} and Eq.~\eqref{eq:sfdsdfsf44}]. Dashed blue line: the weak driving approximation for energy [cf. Eq.~\eqref{eq:sfdsfdfdf}]. Dotted red line: the weak driving approximation for power [cf. Eq.~\eqref{eq:zvxxczxczcx}]. Panel (c): the maximum energy $E(t_E)$ (thick cyan line) and the peak power $P(t_P)$ (thin orange line) that can be achieved in the nonlinear quantum battery [cf. Eq.~\eqref{eq:sfdsdfsf22} and Eq.~\eqref{eq:sfdsdfsf44}]. Dashed blue line: the weak driving approximation for energy [cf. Eq.~\eqref{eq:sfdsfdfdffkuuuf}]. Dotted red line: the weak driving approximation for power [cf. Eq.~\eqref{eq:gfgdbdg}]. Upper row: we consider the damping rate $\gamma = J/2$. Panels (d, e, f): as for panels (a, b, c), but with the loss rate increased fourfold up to $\gamma = 2J$. In this figure we utilize log--log plots and the calculations are performed such the results are convergent up to the energy $10\omega_b$. }
 \label{perff}
\end{figure*}


\subsection{Weak driving limit}
\label{eq:sfdvsdffveecaaa}

Despite the harmful effects of dissipation, it is interesting to consider how the results of the previous subsection are degraded when loss is added to the analysis, since it may be of experimental relevance. Under nonzero damping ($\gamma \ne 0$), the nature of the dynamics of the nonlinear quantum battery changes significantly compared to that encountered in Sec.~\ref{eq:fevtnyny}. For example, the competition between the driving and the dissipation leads to the formation of a nonzero steady state population in the quantum battery. Remaining in the weak driving limit $\Omega \ll J$, the energy $E$ stored in steady state presents the quadratic scaling
\begin{equation}
\label{eq:sfdssdsdffsdf}
\lim_{t \to \infty} E \simeq \omega_b \left( \frac{\Omega}{J} \right)^2,
\end{equation}
which comes directly from the cumulant expansion result of Eq.~\eqref{eq:sssdfsfd}. The system of equations defined by Eqs.~\eqref{eq:sddfsfhhfsdf}--\eqref{eq:sfdfvfvvggg} can be approximately solved analytically within the weak driving regime, as shown explicitly in the Supplemental Material~\cite{Supplemental}. Such a perturbative analysis leads to the following expression for the dynamical energy $E$ stored in the battery
\begin{align}
\label{eq:ssssdfsdfsdfdfsfd}
 E &= \omega_b \left( \tfrac{\Omega}{J} \right)^2 \Bigl\{ 1 - 2 F \mathrm{e}^{-\frac{\gamma t}{4}} + G \mathrm{e}^{-\frac{\gamma t}{2}} \Bigl\}, \\
F &= \cos \left( K t \right) + \tfrac{\gamma}{4 K} \sin \left( K t \right), \\
G &= \tfrac{J^2}{K^2} + \left( 1 - \tfrac{J^2}{K^2} \right) \cos \left( 2 K t \right) + \tfrac{\gamma}{4 K} \sin \left( 2 K t \right),
\end{align}
where the auxiliary and time-dependent functions $F$ and $G$ are defined in terms of the characteristic frequency $K$, where
\begin{equation}
\label{eq:ssssdfsdsfsdffsdfdfsfd}
K = \sqrt{ 2 J^2 - \left( \tfrac{\gamma}{4} \right)^2 }.
\end{equation}
The damped oscillations inherent to Eq.~\eqref{eq:ssssdfsdfsdfdfsfd} are plotted in Fig.~\ref{nonlinny}~(b) as the dashed cyan line for the case where the damping rate $\gamma = J/2$, as well as the numerically exact result (solid blue line), displaying the decay towards the steady state energy (grey line) of Eq.~\eqref{eq:sfdssdsdffsdf}. When the dissipation is not too strong, the optimal charging time $t_E$ and maximum possible energy $E \left( t_E \right)$ are approximately described by [cf. Eq.~\eqref{eq:sfdsdfsf22}]
\begin{align}
\label{eq:sfdsfdfdf}
 t_E &\simeq \frac{\pi}{K}, \\
E \left( t_E \right) &\simeq \omega_b \left( \frac{\Omega}{J} \right)^2 \left( 1 + \mathrm{e}^{-\frac{\pi \gamma}{4 K}} \right)^2, \label{eq:sfdsfdfdffkuuuf}
\end{align}
which are marked with the pink vertical line and orange horizontal line in Fig.~\ref{nonlinny}~(b), which correspond to the dimensionless estimates $J t_E \simeq 2.23$ and $E(t_E) / \lim_{t \to \infty} \{ E \} \simeq 3.27$ respectively in this panel.

The charging power $P$ is likewise plotted underneath in Fig.~\ref{nonlinny}~(e), with the exact result denoted by the solid light-green line and the approximate result by the dashed dark-green line [cf. Eq.~\eqref{eq:sfdsdfsf33} with Eq.~\eqref{eq:ssssdfsdfsdfdfsfd}]. The time of maximal power $t_P$ is roughly given by the dissipationless result of Eq.~\eqref{eq:zvxxczxczcx}, which leads to the approximate maximum charging power
\begin{equation}
\label{eq:gfgdbdg}
 P \left( t_P \right) \simeq \omega_b \frac{\Omega^2}{J} \frac{1}{\sqrt{2} \alpha} \left( 1 - \cos \left(2  \alpha \right) \mathrm{e}^{- \frac{\alpha}{2\sqrt{2}} \frac{\gamma}{J} } \right)^2, 
\end{equation}
where $\alpha = 1.393 \ldots$ is the dimensionless number also appearing in Eq.~\eqref{eq:zvxxczxczcx} and Eq.~\eqref{eq:zvxxczxczcx2}, which is further discussed around there. Hence, within the weak driving limit and with or without losses from the charger accounted for, the charging performance of the nonlinear quantum battery can be quantified at basically the same level of analytic description as was the linear quantum battery.

\subsection{Moderate driving}

Away from the weak driving regime, where the cases without dissipation and with dissipation were treated in Sec.~\ref{eq:fevtnyny} and Sec.~\ref{eq:sfdvsdffveecaaa} respectively (leading to the results presented in the first and second columns of Fig.~\ref{nonlinny}), the quantum master equation of Eq.~\eqref{eq:sjhgfsdasdasdfsfddfsdf} can be readily solved numerically for the mean battery population $\langle b^\dagger b  \rangle$ -- and indeed for other relevant quantities -- using for example the QuTiP quantum toolbox in Python~\cite{Johansson2012}. The results are presented in the semi-logarithmic plots of the final column of Fig.~\ref{nonlinny}, for both a weak driving case (cyan line) and for two successively higher driving amplitudes (orange and pink lines). The time-dependence of the stored energy $E$ is plotted in Fig.~\ref{nonlinny}~(c), highlighting the benefits of having larger drive amplitudes $\Omega$ for both (i) generating a higher energetic peak $E(t_E)$ at some instant in time $t_E$ and (ii) spawning a more impressive steady state energy $\lim_{t \to \infty}(E)$ at long timescales. The corresponding charging power $P$ is shown in the below panel of Fig.~\ref{nonlinny}~(f), confirming the intuitive behaviour that the peak power $P(t_P)$ occurs with increasingly small charging times $t_P$ for increasingly large drive amplitudes.

The overall performance of the quantum battery with nonlinear coupling is summarised in Fig.~\ref{perff}. The steady state behaviour of the energy $E$ (thick cyan line) and ergotropy $\mathcal{E}$ (thin pink line) is considered in Fig.~\ref{perff}~(a), as a function of the drive amplitude $\Omega$ (in units of the coupling rate $J$, and with the example damping rate $\gamma = J/2$). The weak driving approximation (dotted green line) showcases the quadratic dependence of these steady state quantities following Eq.~\eqref{eq:sfdssdsdffsdf}, which is eventually superseded by the cumulant approximation (dashed orange line) and a deviation from a strict quadratic dependence as encapsulated by Eq.~\eqref{eq:sssdfsfd}. With stronger driving amplitudes still, both the steady state energy $E$ and steady state ergotropy $\mathcal{E}$ exhibit sub-quadratic dependencies. Importantly, these steady state quantities are indistinguishable from each other (within the tolerance of the numeric computation) since in the steady state the quantum state can essentially be described by the first and second moments only.

The dynamic performance of the nonlinear quantum battery is displayed in Fig.~\ref{perff}~(b, c). As already previewed in the final column of of Fig.~\ref{nonlinny} for three specific cases, Fig.~\ref{perff}~(b) shows that the specific charging times $t_E$ and $t_P$ at which one should disconnect the charger from the battery in order to optimize the stored energy (thick cyan line) and charging power (thin orange line) quickly reduce with increasing driving amplitude $\Omega$. In the weak driving limit, the upper bounds of these optimal charging times are approximately given by Eq.~\eqref{eq:sfdsfdfdf} for energy (dashed blue line) and Eq.~\eqref{eq:zvxxczxczcx} for power (dashed red line).

The associated optimal energy $E(t_E)$ and optimal charging power $P(t_P)$ are similarly plotted in Fig.~\ref{perff}~(c) as the thick cyan and thin orange lines respectively. The weak driving regime (dashed blue line) reveals that optimal energy $E(t_E)$ has a quadratic dependence on the driving amplitude $\Omega$, with an enhancement compared to the steady state energy [cf. Eq.~\eqref{eq:sfdssdsdffsdf}] due to the second bracketed term in Eq.~\eqref{eq:sfdsfdfdffkuuuf}. While there is a straightforward quadratic dependency on the driving amplitude $\Omega$ for the optimal power $P(t_P)$ in the weak driving regime (dashed red line) thanks to Eq.~\eqref{eq:gfgdbdg}, a deviation is presented for higher driving amplitudes.

For the avoidance of any doubt, we check what happens when the loss rate is increased fourfold up to $\gamma = 2J$ in the lower panels of Fig.~\ref{perff}, where the panels~(d, e, f) correspond to those panels~(a, b, c) sitting directly above them. The steady state results of panel~(d) present minor (almost imperceptible by the eye) differences which suggests that even moderate dissipation rates do not majorly harm the steady state energetics already reported in panel~(a). The optimal charging times of panel~(e) are noticeably different from those of panel~(b) when the driving amplitude $\Omega$ is smaller, but the overall trend of the curves is rather similar to those of panel~(b). A similar pattern is observed in panel~(f), where the fourfold increase in dissipation changes the quantitative relationships between the optimal energy and the charging power with the driving amplitude, but not the qualitative behaviour of the quantum battery as already shown in panel~(c). Overall, our findings suggest that the proposed nonlinear quantum battery presents an interesting quantum system for accumulating reasonable amounts of energy and ergotropy.


\section{Discussion}
\label{eq:Discussion}

We have discussed a class of two-component quantum batteries built from quantum continuous variables. Some fundamental results for Gaussian states lead to the claim that all of the energy stored in such a single-mode bosonic battery can be extracted if the system is also in a minimum uncertainty state. We have explored the consequences of this thermodynamic rule for two example cases, one with linear charger--battery coupling (where this rule applies universally) and the other with a nonlinear charger--battery interaction (where the rule can be applied within certain coupling regimes). In each situation we have judged the quantum energetic performance of the battery in its charging phase, which is mostly describable with short and insightful analytic expressions. 

The more captivating case of the nonlinear quantum battery was inspired by degenerate parametric downconversion, where a photon of frequency $2\omega$ is converted into two photons (each of frequency $\omega$)~\cite{Guo2017, Sandbo2020}. Thanks to the phenomenon of quantum squeezing, a nontrivial quantum state of minimum uncertainty is able to be reached within some parameter regimes. This gives rise to the realization of a plausible energy storage device exploiting a decidedly non-classical phenomenon, as well as one exhibiting an interesting coupling nonlinearity.

We anticipate that our groundwork results should inspire further and more sophisticated experimental and theoretical work on the (perhaps underappreciated) topic of bosonic quantum batteries, especially as the wider field of quantum energetics continues to develop at a rather impressive pace~\cite{Camposeo2025, Gomes2024}.


\section*{Acknowledgments}

\noindent\textit{Funding}: CAD is supported by the Royal Society via a University Research Fellowship (URF\slash R1\slash 201158). \textit{Discussions}: We thank Alan Costa dos Santos (Instituto de Física Fundamental -- CSIC, Madrid) for fruitful discussions as part of their Royal Society International Exchanges grant (IES\slash R1\slash251033) with CAD. \textit{Data and materials availability}: All data is available in the manuscript and the Supplemental Material.
\\



\end{document}